\documentclass[A4,12pt]{article}
\usepackage{jcappub}

\begin{document}
\title {Observational constraints on Modified Chaplygin Gas  from Large Scale Structure}

\author[a]{B. C. Paul,} 
\author[b]{P. Thakur,}
\author[c]{A. Beesham}

\affiliation[a]{Physics Department, North Bengal University \\
 Dist. : Darjeeling, Pin : 734013, West Bengal, India}

\affiliation[b]{Physics Department, Alipurduar College \\
 Dist. : Jalpaiguri,  Pin : 736122, West Bengal, India}

\affiliation[c]{Department of Mathematical Sciences, University of Zululand \\
Private Bag X1001 Kwa-Dlangezwa 3886 South Africa}

\emailAdd{bcpaul@iucaa.ernet.in} 
\emailAdd{abeesham@yahoo.com}

\abstract
{We study cosmological models with  modified Chaplygin gas (in short, MCG) to determine observational constraints on its EoS parameters. The observational data of the background and the growth tests are employed. The background test data namely, $H(z)-z$ data, CMB shift parameter,  Baryonic acoustic oscillations (BAO) peak parameter, SN Ia data  are considered to study the dynamical aspects  of the universe. The growth test data we employ here consists of the linear growth  function for the large scale structures of the universe, models are explored assuming  MCG as a candidate for dark energy. Considering the  observational growth  data  for a given range of redshift from the Wiggle-Z measurements  and rms mass fluctuations from Ly-$\alpha$ measurements, cosmological  models are analyzed numerically to determine constraints on the MCG parameters. In this case, the Wang-Steinhardt ansatz for the  growth index $\gamma$ and growth function $f$ (defined as $f=\Omega_{m}^{\gamma} (a)$) are also taken into account  for the numerical analysis. The best-fit values of the equation of state parameters obtained here are used to study  the  variation of the  growth function ($f$), growth index ($\gamma$), equation of state parameter ($\omega$), squared sound speed $c^2_{s}$ and deceleration parameter with redshift $z$. The observational constraints on the MCG parameters obtained here are then compared  with those of the GCG model for viable cosmology. It is noted that MCG models satisfactorily accommodate an accelerating phase followed by a matter dominated phase of the universe. The permitted range of values of the EoS parameters and the associated parameters ($f$,  $\gamma$, $\omega$, $\Omega$, $c^2_{s}$, $q$) are compared with those obtained earlier using other observations.}


\maketitle
\flushbottom

\section{Introduction}
\label{sec:introd}
Recent  cosmological observations  from supernova  \cite{mut,sb,sn1,sn2,sn4}, WMAP \cite{wmap,wmap2,hin,kog,sper}, BAO  data \cite{bao05}  predicted that the present universe is passing through a phase of accelerating expansion which might be fuelled due to the existence of a new source of energy which is termed as dark energy. In  observational cosmology, the expansion rate $H(z)$ is measured at various redshifts which are useful  to obtain different cosmological  parameters {\it namely}, distance modulus parameter, deceleration parameter. Although the analysis provides us with a satisfactory understanding of cosmological dynamics, it fails to give a complete understanding of the evolution of the universe. Consequently, additional observational input, namely, cosmic growth of the inhomogeneous parts of the universe for its structure formation, is considered in recent times for observational analysis. The growth of the large scale structures  derived from the  linear matter density contrast $\delta(z)\equiv\frac{\delta\rho_{m}}{\rho_{m}}$ of the universe is considered as an  important tool in constraining cosmological model parameters. In order to describe the evolution of the inhomogeneous energy density, it is preferable to parametrize the growth function $f=\frac{d\log\delta}{d\log a}$ in terms of the growth index $\gamma $. It was Peebles \cite{peebles} who first initiated, and thereafter  Wang and Steinhardt \cite{wangstein} parametrized $\delta$ in terms of $\gamma$ to obtain cosmologies which are useful in  various contexts reported in the literature \cite{linder,las,ja,hu,lue,ac,lue3, koivisto,dan,so}. The study  of dark energy for understanding accelerating universe in cosmology  is thus important which may be analyzed using  observational data from the observed  expansion  rate $H(z)$ and growth of matter density contrast $\delta(z)$ data simultaneously.\\
It is known that in the general theory of relativity, ordinary matter fields available from the standard model of particle physics fails to account for the present observations of the universe. It is therefore essential to consider a new type of matter in the  modified sector of matter in the Einstein-Hilbert action or a new physics in this connction.       
  In the literature, Chaplygin gas (CG) was considered to be one such candidate for dark energy. The equation of state (henceforth, EoS) for CG  is 
\begin{equation}
\label{cgeos}
p=-\frac{A}{\rho}
\end{equation}
where $A$ is a positive constant.
It may be important to mention here that the initial idea of a CG originated in aerodynamics \cite{chap}.  CG  may be considered as an alternative to quintessence \cite{kamensh}. In the context of string theory, CG emerges from the dynamics of a generalized d-brane in a (d+1,1) space time. It can be described by a complex scalar field which is obtained from a generalized Born-Infeld action. But CG is  ruled out  in cosmology as cosmological models are not consistent with observational data namely, SNIa, BAO, CMB etc. \cite{zhu04,bento03}. Subsequently  the equation of state for CG is generalized to incorporate  different aspects of the observational  universe. The equation of state for generalized Chaplygin gas (in short, GCG) \cite{bilic, bento} is given by
\begin{equation}
\label{gcgeos}
p=-\frac{A}{\rho^\alpha}
\end{equation}
with $ 0 \leq \alpha \leq 1 $. In the above EoS for $\alpha=1$ reduces to Chaplygin gas\cite{chap}. It has two free parameters $A$ and $\alpha$. It is known that GCG is capable of explaining the background dynamics \cite{jcfab} and various other features of a homogeneous isotropic universe satisfactorily.  The features that the GCG corresponds to almost dust ($p =0$) at high density does not agree completely with our universe.
It is also known that the model suffers from a serious problem  at the perturbative level. The matter power spectrum of GCG exhibits strong oscillations or instabilities, unless GCG model reduces to $\Lambda$CDM \cite{sandvik}. The oscillations for the baryonic component with GCG leads to undesirable features in CMB spectrum \cite{amendola}.\\ In order to use the gas equation in a more satisfactory way, a modification to the GCG  is further considered by adding a positive term, linear in density, to the EoS which is  known as  modified Chaplygin gas (in short MCG).
  The equation of state for the MCG is given by:
\begin{equation}
\label{mcgeos}
p=B\rho-\frac{A}{\rho^\alpha}
\end{equation}
where $A$, $B$, $\alpha$ are positive constants
with $0 \leq \alpha \leq 1 $. The above EoS reduces to that of GCG model \cite{bilic,bento} when one sets  $B=0$. A cosmological constant $\Lambda$  emerges by setting  $\alpha = -1$ and $A = 1 + B$. For $A = 0$, eq. (\ref{mcgeos}) reduces to an EoS which describes a perfect fluid with  $\omega = B$, {\it e.g.},  a quintessence model \cite{lxu}. The MCG contains one more free parameter, namely, $B$, over the GCG. It may be pointed out here that the MCG is a single fluid model which unifies dark matter and dark energy. The  MCG model is  suitable for obtaining  constant negative pressure at low density accommodating late acceleration, and a radiation dominated era (with $B = \frac{1}{3}$) at high density. Thus a universe with a MCG may be described starting from the radiation epoch to the epoch dominated by the dark energy consistently.
On the other hand the GCG describes the evolution of the universe from matter dominated to a dark energy dominated regime (as $B=0$). So compared to GCG, the proposed MCG is suitable to describe the  evolution of the universe over a wide range of epoch \cite{debnath}. 
   On the otherhand the distinction between $\Lambda CDM$  and GCG models are very little, GCG is not very much suitable to describe EoS for the dark energy. Another motivation for considering MCG as a dark energy candidate is that the exact  form of the EoS for dark energy is not yet known. The MCG is  an attempt to find something interesting that is not exactly $\Lambda$CDM. Wu {\it et al.}\cite{wu07} studied the dynamics of
the MCG model. Bedran {\it et al.} \cite{bed08} studied the evolution of the temperature function in the presence of a MCG. It is also consistent with perturbative study \cite{costa1} and the spherical collapse problem \cite{debnath1}. In this paper we consider a universe with MCG for cosmological analysis. \\

In this paper observational constraints on EoS parameters of the MCG are determined using different observational data namely, H(z)-z data, the CMB shift parameter, BAO peak parameter, the SNIa data, the growth function and growth index in a FRW universe.  
The growth data given in Table -\ref{tab2}  consists of a number of data points within redshift ranges (0.15 to 3.0) which is  related to the growth function $f$. It may be pointed out here that the observed growth rate corresponds to various projects/surveys, including the latest Wiggle-Z measurements shown in the table. Gupta {\it et al.} \cite{anjan} obtained constraints on  GCG parameters using the above data. Cosmological models dominated by viscous dark fluids are also considered  in Ref.\cite{velt} where it is shown that viscous fluid mimics a $\Lambda$CDM model when the coefficient of viscosity varies as $\rho^{-1/2}$. It also provides excellent agreement with both the supernova and ($H-z$) data. The viscous cosmological model is analogous to the GCG model. In addition to the above data, other observational sets of growth data given in (Table- \ref{tab3}) from  various sources such as: the redshift distortion of galaxy power spectra \cite{hawkins1}, root mean square ($\it rms$)  mass fluctuation $\sigma_{8}(z)$ obtained from galaxy and $\it Ly$-$\alpha$ surveys at various redshifts \cite{viel1,viel2}, weak lensing statistics \cite{kaiser}, baryon acoustic oscillations \cite{bao05}, X-ray luminous galaxy clusters \cite{manz}, Integrated Sachs-Wolfs (ISW) Effect \cite{rees,am,kaiser3,cr,pog} etc. are important, and which will be taken up here. It is known that redshift distortions are caused by velocity flow induced by gravitational potential gradients which evolved  due to the growth of the universe under gravitational attraction  and dilution of the potentials due to the cosmic expansion. The gravitational growth index $\gamma$ is  considered to be an important parameter which affects the redshift distortion \cite{linder}. The  cluster abundance evolution,  however, strongly depends on $\it rms$ mass fluctuations $\sigma_{8}(z)$ \cite{wangstein} which will be also considered here.\\

We  adopt here $\it chi$-square minimization techniques to constrain different parameters of the EoS for a viable cosmological model considering a MCG as the fluid in the universe. In the analysis total $\it chi$-square is constituted with the background tests  ((i) H(z)-z data (ii)The CMB shift parameter (iii) Baryonic acoustic oscillations (BAO) peak parameter (iv) The SN Ia data) and the growth tests. The best-fit values of the model parameters are then determined from the $\it chi$-square function to study the evolution of the universe.

The paper is organized as follows : 
In sec.2, relevant field equations obtained from the Einstein field equation are given. In sec.3, we determine constraints on the EoS parameters from background test. In sec.4, numerical analysis of the growth index parametrization in terms of the EoS parameters is given defining respective $\it chi$-square functions. In sec.5, constraints on the EoS parameters obtained from the background test and growth test are presented. In sec.6, a summary of the numerical analysis is presented. Finally, in sec.7, we give a brief discussion.       

\section{Einstein Field Equations}

The Einstein field equation is given by
\begin{equation}
\label{ricci}
R_{\mu \nu}-\frac{1}{2} g_{\mu \nu} R = 8 \pi G \; T_{\mu \nu}
\end{equation}
where $R_{\mu \nu}$, $R$, $g_{\mu \nu}$ and $T_{\mu \nu}$ represent the Ricci tensor, the Ricci scalar, the metric tensor in 4-dimensions and the energy momentum tensor respectively.   
We consider a Robertson-Walker  metric which is given by
\begin{equation}
\label{metric}
ds^{2} = - dt^{2} + a^{2}(t) \left[ \frac{dr^{2}}{1- k r^2} + r^2 ( d\theta^{2} + sin^{2} \theta \;
d  \phi^{2} ) \right]
\end{equation}
where  $k=0,+1,-1$ represents flat, closed and open universe, and $a(t)$ is the scale factor of the universe with $r,\theta,\phi$ co-moving co-ordinates.

Using metric (\ref{metric}) in the Einstein field eq. (\ref{ricci}), we obtain the following equations:
\begin{equation}
\label{fried}
3 \left( \frac{\dot{a}^2}{a^2} + \frac{k}{a^2} \right) = 8 \pi G \; \rho, 
\end{equation}
\begin{equation}
2 \frac{\ddot{a}}{a} + \frac{\dot{a}^2}{a^2} + \frac{k}{a^2} = - 8 \pi G \; p,
\end{equation}
where $\rho$ and $p$ represent the energy density and pressure respectively. The conservation equation is given by
\begin{equation}
\label{energy}
\frac{d\rho}{dt} + 3 H \left(\rho + p \right) = 0, 
\end{equation}
where $H = \frac{\dot{a}}{a}$ is Hubble parameter.

Using the EoS given by eq.(\ref{mcgeos}) in eq.(\ref{energy})and integrating once, one obtains the  energy density for a modified Chaplygin gas which is given by 
\begin{equation}
\label{rhomcg}
\rho_{mcg}=\rho_{0}\left[A_{s}+\frac{1-A_{s}}{a^{3(1+B)(1+\alpha)}}\right]^{\frac{1}{1+
\alpha}}
\end{equation}
 where $ A_{s} = \frac{A}{1+B}\frac{1}{\rho_{0}^{\alpha +1}}$ with $ B\neq -1$, $\rho_o$ is an integration constant. The scale factor of the universe is related to the redshift parameter $z$ as $\frac{a}{a_{0}}=\frac{1}{1+z}$,  one may  choose the present scale factor of the universe   $a_{0}=1$ for convenience. The MCG model parameters are  
$ A_{s}$, $ B$  and $\alpha$.
 From eq. (\ref{rhomcg}), it is evident that the positivity condition for the energy density is ensured when $0\leq A_{s} \leq 1$. From  eq. (\ref{rhomcg}), one recovers  the standard $\Lambda$CDM model for  $\alpha= 0 $ and $B = 0$.
The Hubble parameter can be expressed as  a function of  redshift  using the field  eq. (\ref{fried}), which is given by 
\[
H(z)=H_0[\Omega_{b0}(1+z)^3
+ 
\] 
\begin{eqnarray}
\label{hpara}
 \; \; \; (1-\Omega_{b0})
[(A_{s}+(1-A_{s})(1+z)^{3(1+B)(1+\alpha)})^{\frac{1}{1+\alpha}}]]^{\frac{1}{2}}.
\end{eqnarray}
where $\Omega_{b0}$, $H_{0}$ represent the present baryon density and present Hubble parameter, respectively.

The square of the sound speed is given by
\begin{equation}
\label{sound}
c^2_{s}=\frac{\delta p}{\delta\rho}=\frac{\dot{p}}{\dot{\rho}}
\end{equation}
which reduces to
\begin{equation}
\label{sound1}
c^2_{s}=B+\frac{A_{s}\alpha(1+B)}{\left[A_{s}+(1-A_{s})(1+z)^{3(1+B)(1+\alpha)})\right]}.
\end{equation}
In terms of the equation of state  it becomes
\begin{equation}
\label{sound2}
c^2_{s}=-\alpha\omega+B(1+\alpha).
\end{equation}
It may be mentioned here that for causality and stability under perturbations, it is necessary to satisfy  the inequality condition $c_{s}^2 \leq 1$ \cite{lxu}.
The deceleration parameter is given by
\begin{equation}
q(a)=\frac{\frac{\Omega_{b0}}{a^3}+\Omega_{mcg}(a)[1+3\omega(a)]}{2[\frac{\Omega_{b0}}{a^3}+\Omega_{mcg}(a)]}
\end{equation}
where
\begin{equation}
\Omega_{mcg}(a)=\Omega_{mcg0}[A_{s}+\frac{(1- A_{s})}{a^{3(B+1)(1+\alpha)}}]^{\frac{1}{1+\alpha}}
\end{equation}

\section{Background tests from Observed Data}
 
We consider the following  background tests from observed cosmological data for  analyzing cosmological models:
\\
 $\bullet$  The differential age of old galaxies, given by $ H(z)$.\\
 $\bullet$  The peak position of the Baryonic Acoustic Oscillations (BAO).\\
$\bullet$  The CMB shift parameter.\\
$\bullet$  The SN Ia data.\\

The EoS for the MCG contains three unknown parameters namely $A_{s}$, $B$ and $\alpha$, which are  determined here  by a numerical analysis employing different observed data. For the analysis, the Einstein field equation is  rewritten in terms of the Hubble parameter, and a  $\it chi$-square function is also defined corresponding to the observation under consideration.
 
\subsection{$\chi^{2}$-function for Observed Hubble Data (OHD)}
The observed Hubble Data is  given in  table (\ref{tab1}) \cite{stern10} are employed here:
We define  first the $\it chi$-square ($\chi^2_{H-z}$) function  which is 
\begin{equation}
\chi^{2}_{H-z}(H_{0},A_{s}, B,\alpha, z)=\sum\frac{(H(H_{0}, A_{s}, B,\alpha,z)-H_{obs}(z))^2}{\sigma^{2}_{z}}
\end{equation} 
where $H_{obs}(z)$ is the observed Hubble parameter at red shift $z$ and $\sigma_{z}$ corresponds to the error associated with that particular observation  as shown  in table -\ref{tab1}. 
\begin{table}
  \centering
  \begin{tabular}{|l|r|c|c|}
  \hline
  {\it z } & $H(z)$ & $\sigma$ \\
  \hline
   0.00 & 73  & $ \pm $ 8.0	 \\
   0.10 & 69  & $ \pm $ 12.0 \\
   0.17 & 83  & $ \pm $ 8.0 \\
   0.27 & 77  & $ \pm $ 14.0 \\
   0.40 & 95  & $ \pm $ 17.4 \\
   0.48 & 90  & $ \pm $ 60.0 \\
   0.88 & 97  & $ \pm $ 40.4 \\
   0.90 & 117 & $ \pm $ 23.0 \\
   1.30 & 168 & $ \pm $ 17.4 \\
   1.43 & 177 & $ \pm $ 18.2 \\
   1.53 & 140 & $ \pm $ 14.0 \\
   1.75 & 202 & $ \pm $ 40.4 \\

\hline
\end{tabular}
\caption{\label{tab1} $H(z) vs. z$ data from Stern {\it et al.} \cite{stern10}}
\end{table}

\subsection{$\chi^{2}$-function for  Baryon Acoustic Oscillation (BAO)}
 A model independent BAO peak parameter  for low red shift $z_{1}$ measurements in a flat universe is given by \cite{bao05}:
\begin{equation}
\label{baop20}
\mathcal {A} =\frac{\sqrt{\Omega_{m}}}{E(z_{1})^{1/3}}\left(\frac{\int ^{z_{1}}_0 \frac{dz}{E(z)}}{z_{1}}\right)^{2/3} 
\end{equation}
where $\Omega_{m}$ is the matter density parameter for the Universe. The $\it chi$-square function in this case is  defined as :
\begin{equation}
\label{chibao20}
\chi^{2}_{BAO}(A_{s},B,\alpha,z)=\frac{\left(\mathcal{A}-0.469\right)^{2}}{ \left(0.017\right)^{2}}.
\end{equation}
The Sloan Digital Sky Survey (SDSS) data for the Luminous Red Galaxies  (LRG) survey gives  $\mathcal{A}$ ($0.469\pm.0.017$)  \cite{bao05}.

\subsection{$\chi^{2}$-function for Cosmic Microwave Background (CMB)}
The CMB shift parameter ($\mathcal {R}$) is given by \cite{komat11}:
\begin{equation}
\label{cmbp20}
\mathcal{R}=\sqrt{\Omega_{m}}\int ^{z_{ls}}_{0} \frac{dz'}{H(z')/H_{0}}
\end{equation}
where $z_{ls}$ is the value of $z$ at the surface of last scattering. The WMAP7 data predicts $\mathcal{R}=1.726 \pm 0.018$ at $z=1091.3$. We now define $\it chi$-square function  as :
\begin{equation}
\label{chicmb20}
\chi^2_{CMB}(A_{s},B,\alpha,z)=\frac{(\mathcal{R}-1.726)^2}{(0.018)^2}.
\end{equation}

Now we combine the above three $\it chi$-square functions as  $\chi ^2_{hbc} = \chi ^2_{H-z}+\chi ^2_{BAO}+\chi ^2_{CMB}$. 
\begin{figure}
\centering
{\includegraphics[width=8cm,height=6cm]{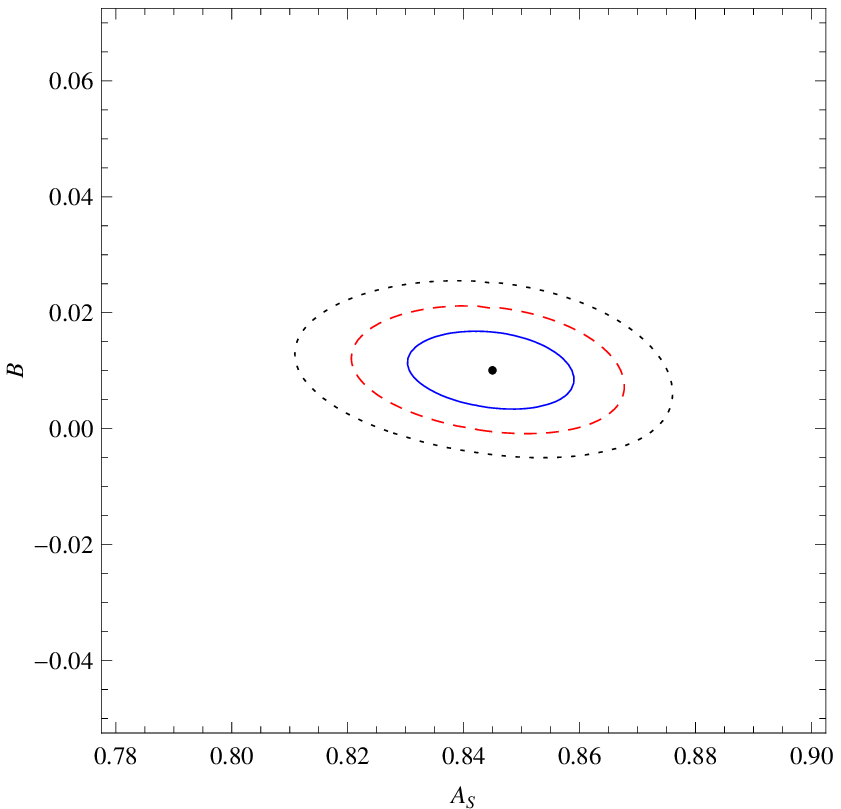}}\\
{\includegraphics[width=8cm,height=5cm]{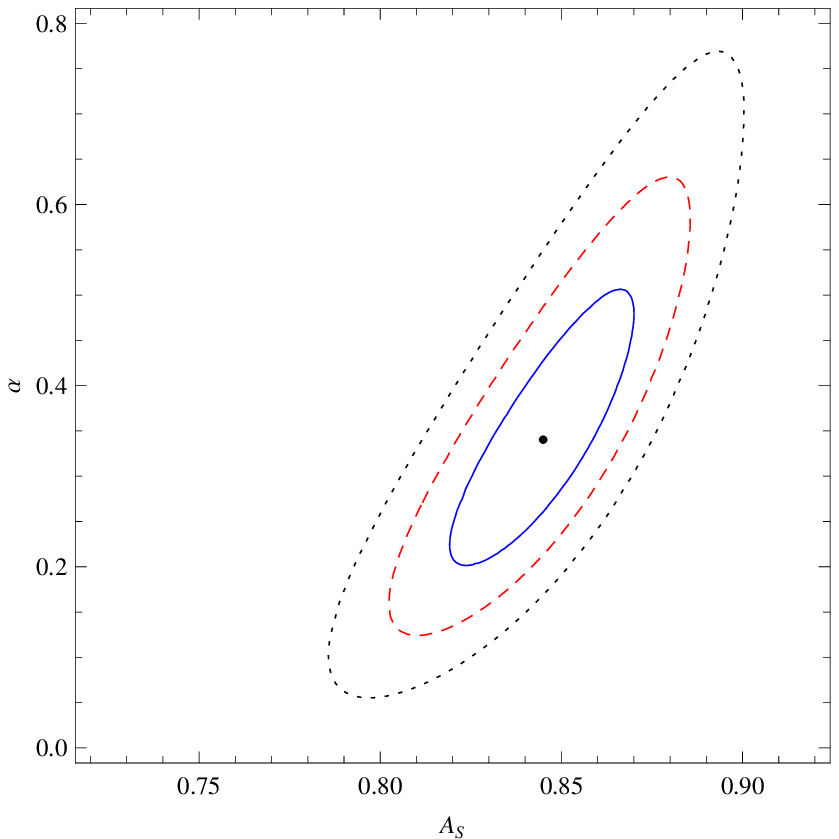}}\\
{\includegraphics[width=8cm,height=5cm]{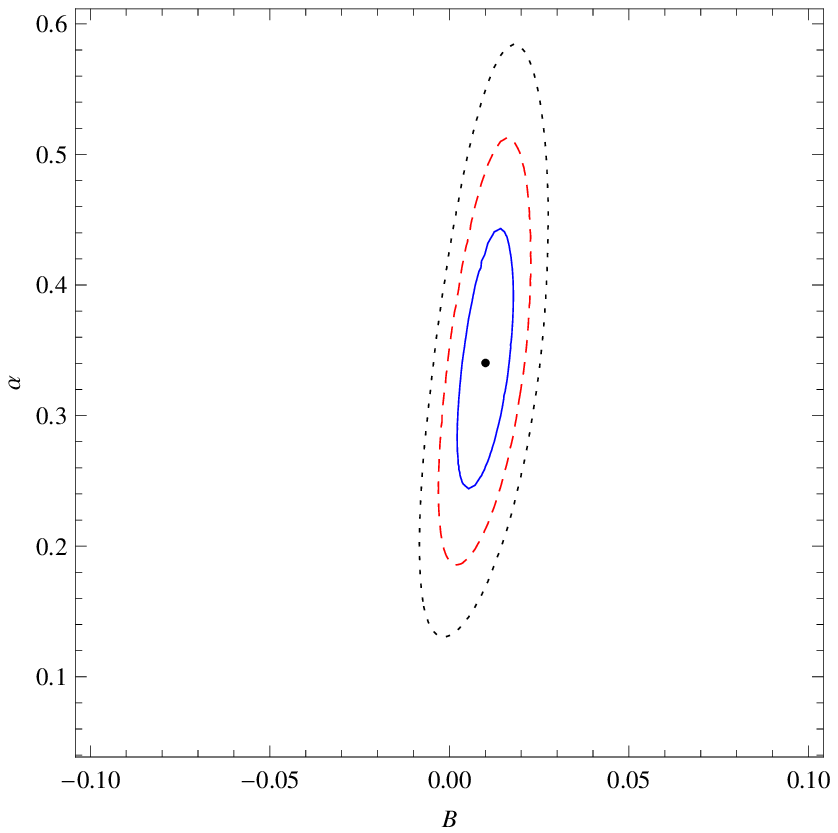}}\\
\caption{Contours  from(i) (H-z)+BAO+CMB+Union2 supernovae data at  $68.3$\%(Solid) $95.4\%$ (Dashed) and $99.7 \%$  (Dotted) confidence limit }
\label{back21}
\end{figure}

\subsection{$\chi^{2}$-functions for Supernovae Data}
The distance modulus function ($\mu$) is defined  in terms of the luminosity distance ($d_{L}$)  as
\begin{equation}
\label{mu10}
\mu(A_{s},B,\alpha,z)= m-M = 5\log_{10}(d_{L})+25
\end{equation}
where 
\begin{equation}
\label{lum10}
d_{L}=\frac{c(1+z)}{H_{0}} \int ^{z}_{0} \frac{dz'}{E(z')}
\end{equation}

In this case the $\it chi$-square ($\chi^2_{\mu}$) function  is defined as
\begin{equation}
\chi^{2}_{\mu}(A_{s},B,\alpha,z)=\sum\frac{(\mu(A_{s},B,\alpha,z)-\mu_{obs}(z))^2}{\sigma^{2}_{z}}
\end{equation} 
where $\mu_{obs}(z)$ is the observed distance modulus at red shift $z$ and $\sigma_z$ is the corresponding error for the observed data\cite{wang11}.
\subsection{Combined $\chi^{2}$-function for the Background Tests}
Finally the total $\chi^{2}$-function for background tests is defined as follows
\begin{equation}
\label{chiback}
\chi^{2}_{back}(A_{s},B,\alpha,z)=\chi^2_{H-z}(A_{s},B,\alpha,z) +\chi ^2_{BAO}(A_{s},B,\alpha,z)+\chi ^2_{CMB}(A_{s},B,\alpha,z)+\chi^{2}_{\mu}(A_{s},B,\alpha,z)
\end{equation}

\begin{figure}
\centering
{\includegraphics[width=8cm,height=6cm]{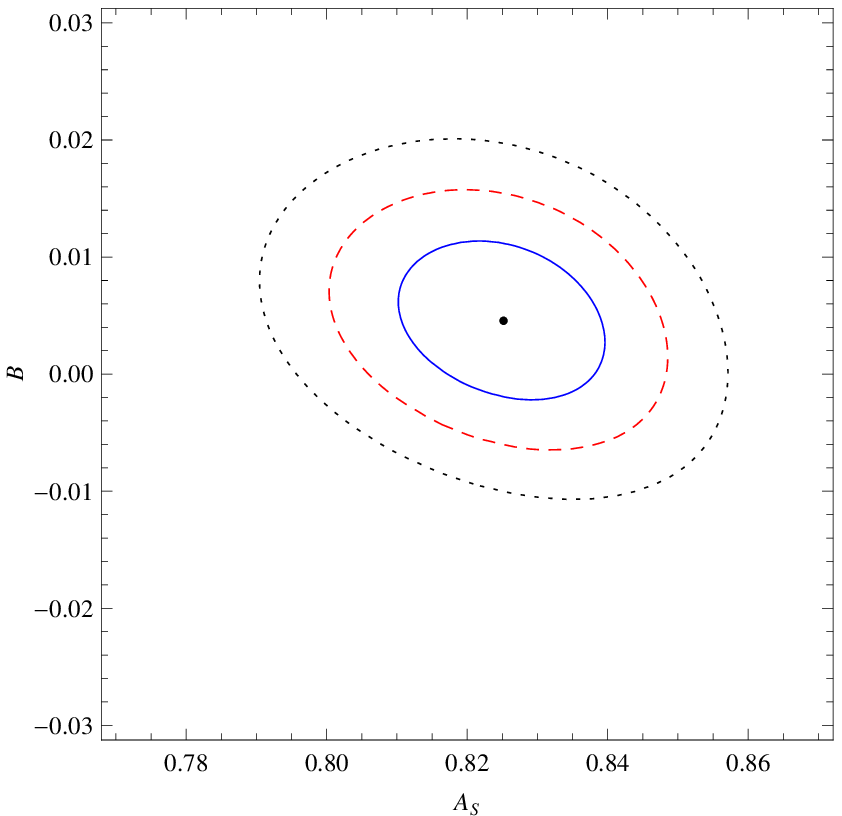}}\\
{\includegraphics[width=8cm,height=5cm]{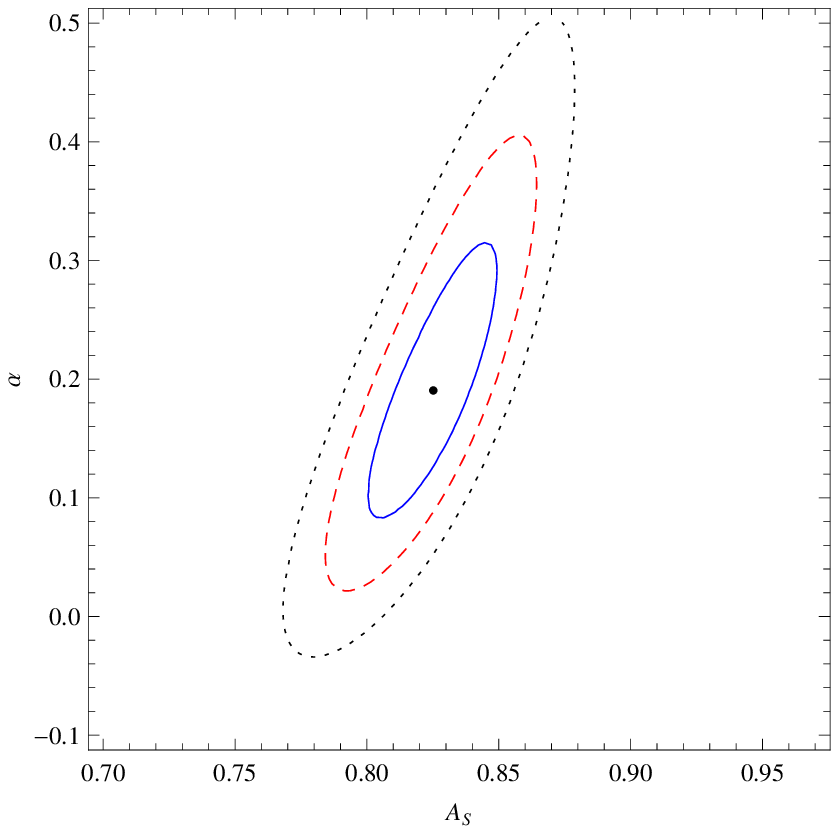}}\\
{\includegraphics[width=8cm,height=5cm]{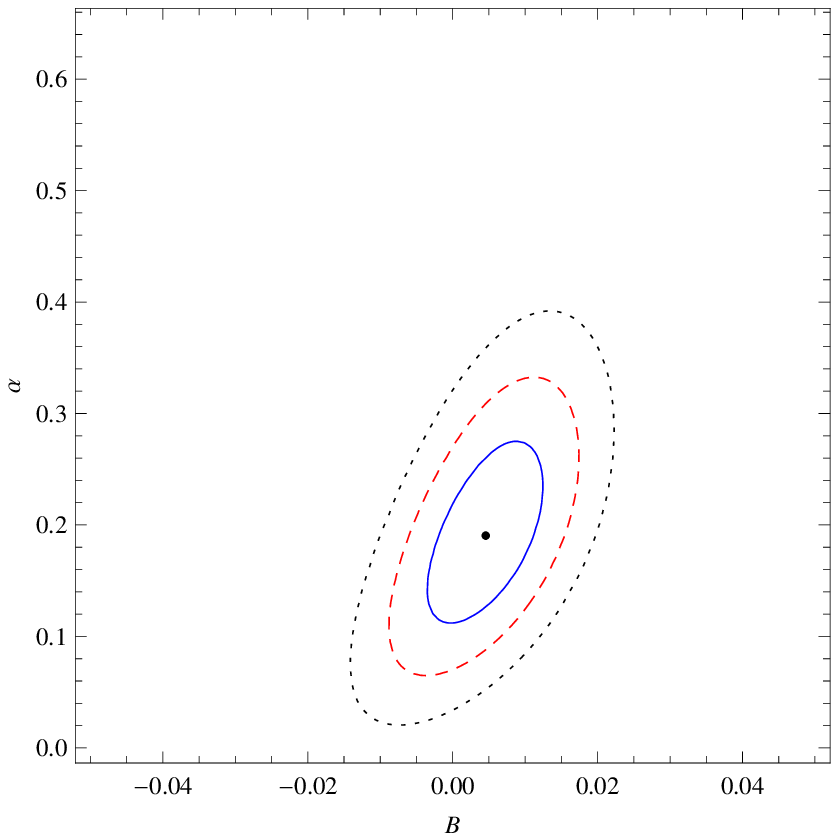}}\\
\caption{Contours  from(i) (H-z)+BAO+CMB+Union2+growth+rms mass fluctuation data at  $68.3$\%(Solid) $95.4\%$ (Dashed) and $99.7 \%$  (Dotted) confidence limit }
\label{tot21}
\end{figure}
The  $\chi^{2}$ function for the background test is minimized by the present Hubble value predicted by WMAP7. The best fit values of $A_{s}$, $B$, $\alpha$ are thereafter determined. 

\section{ Parametrization of the Growth Index}

 The growth rate of large scale structures is derived from the matter density perturbation $\delta=\frac{\delta\rho_{m}}{\rho_{m}}$ (where  $\delta\rho_{m}$ represents the fluctuation of matter density $\rho_{m}$) in the linear regime which satisfies
\begin{equation}
\label{grrate}
\ddot{\delta}+2\frac{\dot{a}}{a}\dot{\delta}-4\pi G_{eff}\rho_{m}\delta=0.
\end{equation}
The field equation for the background cosmology comprising both  matter and MCG in FRW universe are given below
\begin{equation}
\left(\frac{\dot{a}}{a}\right)^{2}=\frac{8\pi G}{3}(\rho_{b}+\rho_{mcg}),
\end{equation}
\begin{equation}
2\frac{\ddot{a}}{a}+\left(\frac{\dot{a}}{a}\right)^{2}=-8\pi G\omega_{mcg}\rho_{mcg}
\end{equation}
where $\rho_{b}$ represents the background energy density and $\omega_{mcg}$  represents the equation of state for the MCG which is given by
\begin{equation}
\omega_{mcg}=B-\frac{A_{s}(1+B)}{\left[A_{s}+(1-A_{s})(1+z)^{3(1+B)(1+\alpha)})\right]}.
\end{equation}
 We now replace the time $t$ variable in terms of $\ln a$ in  eq.(\ref{grrate}) and obtain
\begin{equation}
\label{delta}
(\ln\delta)^{''}+(\ln\delta)^{'2}+(\ln\delta)^{'}
\left[\frac{1}{2}-\frac{3}{2}\omega_{mcg}(1-\Omega_{m}(a))\right]=\frac{3}{2}\Omega_{m}(a)
\end{equation}
where
\begin{equation}
\label{density}
\Omega_{m}(a)=\frac{\rho_{m}}{\rho_{m}+\rho_{mcg}}.
\end{equation}
The  effective matter density is given by  $\Omega_{m}=\Omega_{b}+(1-\Omega_{b})(1-A_{s})^{(1/1+\alpha)}$ \cite{li}.
Using the energy conservation eq. (\ref{energy}) and changing the variable from ${\it ln} a$ to $\Omega_{m}(a)$ once again,  the eq. (\ref{delta}) can be expressed in terms of the logarithmic growth factor $f=\frac{d\log\delta}{d\log a}$ which is given by 
\begin{equation}
\label{gfactor}
3\omega_{mcg}\Omega_{m}(1-\Omega_{m})\frac{d f}{d \Omega_{m}}+f^{2}
 +f\left[\frac{1}{2}-\frac{3}{2}\omega_{mcg}(1-\Omega_{m}(a))\right]=\frac{3}{2}\Omega_{m}(a).
\end{equation}

The logarithmic growth factor $f$, according to Wang and Steinhardt \cite {wangstein}, is given by
\begin{equation}
\label{ansatzf}
f=\Omega_{m}^{\gamma}(a)
\end{equation} 
where $\gamma$ is the growth index parameter. In the case of a flat dark energy model  with constant equation of state  $\omega_{0}$, the growth index $\gamma$ is given by
\begin{equation}
\label{gamflat}
\gamma=\frac{3(\omega_{0}-1)}{6\omega_{0}-5}.
\end{equation}
For a  $\Lambda$CDM model, it reduces to $\frac{6}{11}$ \cite{linder,evl}, for a matter dominated model, one gets $\gamma=\frac{4}{7}$ \cite{fry,ne}. One can also write $\gamma$  as a parametrized function of redshift parameter $z$.  One such parametrization is
$\gamma(z)=\gamma(0)+\gamma^{'}z$, with $\gamma^{'}\equiv\frac{d\gamma}{dz}|_{(z=0)}$ \cite{polarski, gan}. It has been shown recently \cite{ishak} that the parametrization smoothly interpolates a low and intermediate redshift range to a high redshift range up to the cosmic microwave background scale. The above parametrization is also taken up in different contexts \cite{dosset}.
In this paper we parametrize  $\gamma$  in terms of the MCG parameters namely,  $A_{s}$, $\alpha$ and $B$.
 Therefore, we begin with the following  ansatz  given by
\begin{equation}
\label{ansatzmcg}
f=\Omega_{m}^{\gamma(\Omega_{m})} (a)
\end{equation} 
where the growth index parameter $\gamma(\Omega_{m})$ can be expanded in a Taylor series around $\Omega_{m} = 1$ as
\[
\gamma(\Omega_{m}) \hspace{8cm}
\]
\begin{equation}
\label{tylor}
\;\; \; \; \; \; =\gamma|_{(\Omega_{m}=1)}+(\Omega_{m}-1)\frac{d\gamma}{d\Omega_{m}}|_{(\Omega_{m}=1)} +O(\Omega_{m}-1)^{2}.
\end{equation}
Consequently eq.(\ref{gfactor}) can be rewritten in terms of $\gamma$ as
\[
3\omega_{mcg}\Omega_{m}(1-\Omega_{m})\ln\Omega_{m}\frac{d\gamma}{d \Omega_{m}}-3\omega_{mcg}\Omega_{m}(\gamma-\frac{1}{2})+
\]
\begin{equation}
\label{tylomega}
\; \; \; \; \; \Omega_{m}^{\gamma}-\frac{3}{2}\Omega_{m}^{1-\gamma}+3\omega_{mcg}\gamma-\frac{3}{2}\omega_{mcg}+\frac{1}{2}=0.
\end{equation}
Differentiating once again the above equation around $\Omega_{m}=1$, one obtains  zeroth order term in the expansion  for $\gamma$ which is given by
\begin{equation}
\label{grmcg}
\gamma=\frac{3(1-\omega_{mcg})}{5-6\omega_{mcg}}.
\end{equation}
It  agrees with a dark energy model for a constant $\omega_{0}$ (eq. \ref{gamflat}). 
In the same way differentiating it twice, followed thereafter by a  Taylor expansion  around $\Omega_{m}=1$, one obtains the first order terms  in the expansion which is given by
\begin{equation}
\label{forder}
\frac{d\gamma}{d\Omega_{m}}|_{(\Omega_{m}=1)} = \frac{3(1-\omega_{mcg})(1-\frac{3\omega_{mcg}}{2})}{125(1-\frac{6\omega_{mcg}}{5})^{3}}.
\end{equation}
Substituting it in eq. (\ref{tylor}), $\gamma$ is obtained,  the first order term in this case   approximating to
\begin{equation}
\label{gammamcg}
\gamma=\frac{3(1-\omega_{mcg})}{5-6\omega_{mcg}}+(1-\Omega_{m})\frac{3(1-\omega_{mcg})(1-\frac{3\omega_{mcg}}{2})}{125(1-\frac{6\omega_{mcg}}{5})^{3}}.
\end{equation}
Using the expression of $\omega_{mcg}$ in the above,  $\gamma$ can be parametrized with the MCG  parameters.
 We define the normalized growth function $g$ from the numerically obtained solution using eq. (\ref{delta}) which is given by 
 \begin{equation}
 \label{g}
 g(z)\equiv\frac{\delta(z)}{\delta(0)}.
 \end{equation}
 The corresponding approximate  normalized growth function obtained from the parametrized form of $f$  follows from eq.(\ref{ansatzmcg}) and  is given by
 \begin{equation}
 g_{th}(z)=\exp\left( \int_{1}^{\frac{1}{1+z}}  \Omega_{m}(a)^{\gamma}\frac{da}{a} \right).
 \end{equation}
  The above expression will be  employed once again to construct  $\it chi$-square functions in the next section. 
	
 \subsection{$\chi^{2}$ for growth function (f) }

\begin{table}
\centering
		\begin{tabular}{@{}|l|c|c|r|}
		\hline
		z     & $f_{obs}$ &    $\sigma$ &    $Ref.$\\
		\hline
		$0.15$   & 0.51 &  0.11 &  \cite{hawkins,verde}\\
		$0.22$ & 0.60 & 0.10 &  \cite{blake}\\
		$0.32$ & 0.654 & 0.18 & \cite{reyes}\\
		$0.35$ & 0.70 & 0.18 &  \cite{tegmark}\\
		$0.41$ & 0.70 & 0.07 & \cite{blake}\\
		$0.55$ & 0.75 & 0.18 & \cite{ross}\\
		$0.60$ & 0.73 & 0.07 & \cite{blake}\\
		$0.77$ & 0.91 & 0.36 & \cite{guzzo}\\
		$0.78$ & 0.70 & 0.08 & \cite{blake}\\
		$1.4$ & 0.90 & 0.24 & \cite{angela}\\
		$3.0$ & 1.46 & 0.29 & \cite{mcdon}\\
		\hline	
		\end{tabular}
\caption{Data for the observed growth functions $f_{obs}$ used in our analysis }	
	\label{tab2}
\end{table}

\begin{table}
\centering
		\begin{tabular}{@{}|l|c|c|r|}
		\hline
		z     & $\sigma_{8}$ &    $\sigma_{\sigma_{8}}$ &    $Ref$\\
		\hline
		$2.125$   & 0.95 &  0.17 &  \cite{viel1}\\
		$2.72$ & 0.92& 0.17 &       \\
		$2.2$ & 0.92 & 0.16 & \cite{viel2}\\
		$2.4$ & 0.89 & 0.11 &   \\
		$2.6$ & 0.98 & 0.13 &   \\
		$2.8$ & 1.02 & 0.09 & \\
		$3.0$ & 0.94 & 0.08 & \\
		$3.2$ & 0.88 & 0.09 & \\
		$3.4$ & 0.87 & 0.12 & \\
		$3.6$ & 0.95 & 0.16 & \\
		$3.8$ & 0.90 & 0.17 & \\
		$0.35$ & 0.55 & 0.10 & \cite{marin}\\
		$0.6$ & 0.62 & 0.12 &    \\
		$0.8$ & 0.71 & 0.11 &    \\
		$1.0$ & 0.69 & 0.14 &    \\
		$1.2$ & 0.75 & 0.14 &    \\
		$1.65$ & 0.92 & 0.20 &   \\
		\hline	
		\end{tabular}
\caption{Data for the $\it rms$ mass fluctuations ($\sigma_{8}$) at various redshift}	
	\label{tab3}
\end{table}
We define $chi$-square for the growth function $f$ as
\begin{equation}
\chi^{2}_{f}(A_{s},B,\alpha,z)=\Sigma\left[\frac{f_{obs}(z_{i})-f_{th}(z_{i},\gamma)}{\sigma_{f_{obs}}}\right]^{2}
\end{equation}
where $f_{obs}$ and $\sigma_{f_{obs}}$ are obtained from table (\ref{tab2}). However, $f_{th}(z_{i},\gamma)$ is obtained from eqs. (\ref{ansatzmcg}) and (\ref{gammamcg}).
 Another  observational probe for the matter density perturbation $\delta(z)$ is derived from the red shift dependence of the rms mass fluctuation $\sigma_{8}(z)$. The dispersion of the density field $\sigma^{2}(R,z)$ on a co-moving  scale $R$ is defined as
\begin{equation}
\sigma^{2}(R,z)=\int_{0}^{\inf} W^{2}(kR)\Delta^{2}(k,z)\frac{dk}{k}
\end{equation}
where 
\begin{equation}
W(kR)=3\left(\frac{\sin(kR)}{(kR)^{3}}-\frac{\cos(kR)}{(kR)^{2}}\right),
\end{equation}
represents window function, and
\begin{equation}
\Delta^{2}(kz)=4\pi k^{3}P_\delta(k,z),
\end{equation}
where $P_\delta(k,z)\equiv(\delta^{2}_{k})$ is the mass power spectrum at red-shift $z$. The {\it rms}  mass fluctuation $\sigma_{8}(z)$ is the  $\sigma^{2}(R,z)$  at  $R=8h^{-1}$ Mpc. The function $\sigma_{8}(z)$ is  connected to $\delta(z)$ as
\begin{equation}
\sigma_{8}(z)=\frac{\delta(z)}{\delta(0)}\sigma_{8}|_{(z=0)}
\end{equation}
which implies
\begin{equation}
s_{th}(z_{1}, z_{2}) \equiv \frac{\sigma_{8}(z_{1})}{\sigma_{8}(z_{2})}=\frac{\delta(z_{1})}{\delta(z_{2})}=
\frac{\exp\left[ \int_{1}^{\frac{1}{1+z_{1}}}  \Omega_{m}(a)^{\gamma}\frac{da}{a} \right] }{\exp \left[ \int_{1}^{\frac{1}{1+z_{2}}}  \Omega_{m}(a)^{\gamma}
\frac{da}{a} \right]}.
\end{equation}
In tab-\ref{tab3}, a systematic evolution of the rms mass fluctuation $\sigma_{8}(z_{i})$ with observed red shift for flux power spectrum of $\it Ly$-$\alpha$ forest \cite{viel1,viel2,marin} is displayed. In this context we  define  a new chi-square  function which is given by 
\begin{equation}
\chi^{2}_{s}(A_{s},B,\alpha,z)=\Sigma\left[\frac{s_{obs}(z_{i},z_{i+1})-s_{th}(z_{i},z_{i+1})}{\sigma_{s_{obs,i}}}\right]^{2}.
\end{equation}
Data for rms mass fluctuation at various red shift given in table-\ref{tab3} will be considered here.
Now considering the growth function mentioned above, one can  define a $\it chi$-square function which  is given by 
\begin{equation}
\label{chigr}
\chi^{2}_{growth}(A_{s},B,\alpha,z)=\chi^{2}_{f}(A_{s},B,\alpha,z)+\chi^{2}_{s}(A_{s},B,\alpha,z).
\end{equation}
The $\it chi$-square functions defined above will be considered for the analysis in the next section.

\subsection{Combined $\chi^{2}$ function for the background test and growth test}
Using eq.(\ref{chiback}) and eq.(\ref{chigr}),  we define total $\it chi$-square function as
\begin{equation}
\chi^{2}_{total}(A_{s},B,\alpha,z)=\chi^{2}_{back}(A_{s},B,\alpha,z)+\chi^{2}_{growth}(A_{s},B,\alpha,z)
\end{equation}
where
$\chi^{2}_{growth}(A_{s},B,\alpha,z)=\chi^{2}_{f}(A_{s},B,\alpha,z)+\chi^{2}_{s}(A_{s},B,\alpha,z)$.
In this case the best fit values are obtained  minimizing  the $\it chi$-square function. Since the $\it chi$-square function depends on $A_{s}$, $B$, $\alpha$ and $z$, it is possible to draw contours at different confidence limits. The limits imposed by the contours  determines the permitted range of values of the EoS parameters in the MCG model.

\begin{table}[tbp]
\centering
		\begin{tabular}{|l|r|c|c|c|}
		\hline
		Data     & $A_{s}$ &    $B$ &    $\alpha$ &  $\chi^{2}/d.o.f$\\
		\hline
		$OHD+BAO+CMB+Union2$  & 0.8450 &  0.0101 &  0.3403 & 1.0406\\
		$OHD+BAO+CMB+Union2$    &        &       &         &    \\
		$+Growth+\sigma_{8}$ & 0.8252 &0.0046&0.1905 &1.0296 \\
	
	  \hline	
		\end{tabular}
\caption{\label{tab4} Best-fit values of the EoS parameters}		
\end{table}

\begin{table}
\centering
		\begin{tabular}{|l|r|c|c|}
		\hline
		Data &  $CL$  & $A_{s}$ &  $B$ \\
		\hline
		$OHD+BAO+CMB+Union2$  & $68.3\%$ & $(0.8303,\;0.8593)$ & $(0.0034,\; 0.0170)$  \\
		$OHD+BAO+CMB+Union2$  & $95.4\%$ & $(0.8202,\;0.8678)$ & $(-0.0010,\;0.0212) $\\
		$OHD+BAO+CMB+Union2$ & $99.7\%$  & $(0.8105,\;0.8762)$& $(-0.0052,\;0.0257)$\\
	\hline	
		\end{tabular}
\caption{\label{tab5} Range of  values of the EoS parameters $A_{s}$ \&  $B$ from background tests and $OBC=OHD+BAO+CMB$}		
\end{table}
\begin{table}
\centering
		\begin{tabular}{|l|r|c|c|}
		\hline
		Data &  $CL$  & $A_{s}$ &  $B$ \\
		\hline
		$OBCU+Growth+\sigma_{8}$ & $68.3\%$ & $(0.8102,\; 0.8398)$ & $(-0.0023,\; 0.0115)$\\
		$OBCU+Growth+\sigma_{8}$ & $95.4\%$ & $(0.8002,\; 0.8488)$ & $(-0.0065,\; 0.0157)$\\
		$OBCU+Growth+\sigma_{8}$ & $99.7\%$ & $(0.7906,\; 0.8571)$ & $(-0.0107,\; 0.0203)$\\
	\hline	
		\end{tabular}
\caption{\label{tab5a} Range of  values of the EoS parameters $A_{s}$ \&  $B$ from combined tests and $OBCU=OHD+BAO+CMB+Union2$}		
\end{table}
\begin{table}
\centering
		\begin{tabular}{|l|r|c|c|}
		\hline
		Data &  $CL$  & $A_{s}$ & $\alpha$ \\
		\hline
		$OHD+BAO+CMB+Union2$  & $68.3\%$ & $(0.8184,\; 0.871)$ & $(0.1988,\;0.5066)$ \\
		$OHD+BAO+CMB+Union2$  & $95.4\%$ & $(0.8024,\; 0.8858)$ & $(0.1218,\; 0.6271) $\\
		$OHD+BAO+CMB+Union2$  & $99.7\%$ & $(0.7851,\; 0.9005)$ & $(0.0526,\;0.7708)$\\
	  \hline	
		\end{tabular}
\caption{\label{tab6} Range of  values of the EoS parameters $A_{s}$ \& $\alpha$ from background tests}		
	\end{table}
	\begin{table}
\centering
		\begin{tabular}{|l|r|c|c|}
		\hline
		Data &  $CL$  & $A_{s}$ & $\alpha$ \\
		\hline
		$OBCU+Growth+\sigma_{8}$ & $68.3\%$ &  $(0.8005,\; 0.8492)$ & $(0.0821,\;0.3141)$ \\
		$OBCU+Growth+\sigma_{8}$ & $95.4\%$ & $(0.7837, \; 0.8642)$ & $(0.0212,\; 0.4065)$\\
		$OBCU+Growth+\sigma_{8}$ & $99.7\%$ & $(0.7678,\; 0.8793)$ & $(-0.0339,\;0.5048)$\\
	  \hline	
		\end{tabular}
\caption{\label{tab6a} Range of  values of the EoS parameters $A_{s}$ \& $\alpha$ from combined tests, and $OBCU=OHD+BAO+CMB+Union2$}		
	\end{table}
\begin{table}
\centering
		\begin{tabular}{|l|r|c|c|}
		\hline
		Data &  $CL$  & $B$ & $\alpha$ \\
		\hline
		$OHD+BAO+CMB+Union2$  & $68.3\%$ & $(0.0021,\; 0.0175)$ & $(0.2433,\;0.4415)$ \\
		$OHD+BAO+CMB+Union2$  & $95.4\%$ & $(-0.0027,\; 0.0226)$ & $(0.1856,\;0.5107) $\\
		$OHD+BAO+CMB+Union2$ & $99.7\%$  & $(-0.00816,\; 0.0274)$ & $(0.1298,\;0.5857)$\\
		 \hline	
		\end{tabular}
\caption{\label{tab7} Range of  values of the EoS parameters $B$ \& $\alpha$ from background tests }	
	\end{table}
\begin{table}
\centering
		\begin{tabular}{|l|r|c|c|}
		\hline
		Data &  $CL$  & $B$ & $\alpha$ \\
		\hline
		$OBCU+Growth+\sigma_{8}$& $68.3\%$ & $(-0.0034,\; 0.0124)$ & $(0.1110,\;0.2755)$ \\
		$OBCU+Growth+\sigma_{8}$ & $95.4\%$ & $(-0.0086,\;0.0174)$ & $(0.0634,\;0.3332)$\\
		$OBCU+Growth+\sigma_{8}$ & $99.7\%$& $(-0.0143,\; 0.0222)$ & $(0.0187,\;0.3909)$\\
		 \hline	
		\end{tabular}
\caption{\label{tab7a} Range of values of the EoS parameters $B$ \& $\alpha$ from combined tests and $OBCU=OHD+BAO+CMB+Union2$}	
	\end{table}
\begin{table}
\centering
		\begin{tabular}{|l|r|c|c|c|c|c|c|c|}
		\hline
		Model &   $A_{s}$ &  $B$ & $\alpha$ & $f$ & $\gamma$&$\Omega_{m0}$& $\omega_{0}$ & $q(0)$\\
		\hline
		$MCG$ & $0.825$   &  $0.005$ & $0.190$ & $0.472$& $0.559$& $0.261$& $-0.836$ & $-0.691$\\
		$GCG$ & $0.819$   &  $0.0$ & $0.141$ & $0.467$& $0.559$& $0.255$& $-0.822$& $-0.684$\\
		
		$MCG(prev)$ & $0.769$   &  $0.008$ & $0.002$ & $0.472$& $0.562$& $0.262$& $-0.767$& $-0.607$\\
		        
		$GCG(prev)$ & $0.708$   &  $0.0$ & $-0.140$ & $0.477$& $0.564$& $0.269$& $-0.708$& $-0.527$ \\
		                    
		$\Lambda$CDM(prev) &  $0.761$   &  $0.0$ & $0.0$ & $0.479$& $0.562$& $0.269$& $-0.761$& $-0.603$ \\
		                        
	  \hline	
		\end{tabular}
\caption{\label{tab8} Values of the EoS parameters in different model}	
	
\end{table}

\begin{table}
\centering
		\begin{tabular}{@{}|l|c|c|c|c|r|}
		\hline
		Model& Data     & $A_{s}$ &    $\alpha$ & B &   $Ref.$\\
		\hline
		$GCG$ & $Supernovae$   & 0.6-0.85 &  -- & 0.0& \cite{makler}\\
		$GCG$&$CMBR$   & 0.81-0.85 & 0.2-0.6 & 0.0& \cite{bentocmb}\\
		$GCG$& $WMAP$ & 0.78-0.87 & -- & 0.0&\cite{bentowmap}\\
		 $GCG$&$CMBR+BAO$ & $\approx0.77$ & $\leq0.1$ & 0.0& \cite{barr}\\
		   $MCG$&$OBCU+Growth+\sigma_{8}$ & 0.825 & 0.190 & 0.005 & this paper\\
			 $MCG$&$Growth+\sigma_{8}+OHD$ & 0.769 & 0.002 & 0.008 & \cite{paul13}\\
		
		\hline	
		\end{tabular}
\caption{Comparison of the values of EoS parameters for GCG and MCG  models (using $ OBCU= OHD+BAO+CMB+Union2$}	
	\label{tab9}
\end{table}

\section{Summary of the Analysis}

The best-fit values of the  EoS parameters of the MCG model as determined from  the two $\it chi$-squares are shown in Table-\ref{tab4}. 
 Contours are drawn for (i) $B$ vs. $A_{s}$ in figs.\ref{back21}(a) and \ref{tot21}(a), (ii)  $\alpha$ vs. $A_s$ in figs. \ref{back21}(b) and \ref{tot21}(b), and (iii) $\alpha$ vs. $B$ in figs.\ref{back21}(c) and \ref{tot21}(c) respectively. 
The allowed ranges of values of the EoS parameters  are  estimated from the above drawn contours. In Table-\ref{tab5} and \ref{tab5a}  the range of values of $A_{s}$ and $B$ from background test and combined tests at different confidence levels are tabulated. We note that $A_{s}$ lies in the interval  $(0.8105,\;0.8762)$ and $B$ lies in the interval $(-0.0052,\;0.0257)$ at 3$\sigma$ level in the background test. However, in the combined test, $A_{s}$ and $B$ satisfy the ranges $(0.7906,\; 0.8571)$ \& $(-0.0107,\; 0.0203)$, respectively, at 3$\sigma$ level.\\

 In Table-\ref{tab6} and \ref{tab6a}  the allowed ranges of values of $A_{s}$ and $\alpha$ are presented that are obtained from the background test as well as those are obtained from the combined tests at different confidence levels. It is evident that for background test $A_{s}$ and $\alpha$ lie in the intervals $(0.7851,\; 0.9005)$ \& $(0.0526,\;0.7708)$, respectively at 3$\sigma$ level, whereas the intervals are  $(0.7678,\; 0.8793)$ \& $(-0.0339,\;0.5048)$, respectively, at 3$\sigma$ level in the case of the combined test. In Table-\ref{tab7} and \ref{tab7a} the allowed range of values of $\alpha$ and $B$ are tabulated both for background test and combined tests at different confidence levels. Here, $B$ and $\alpha$ lie in the interval  $(-0.00816,\; 0.0274)$ \& $(0.1298,\;0.5857)$, respectively, at 3$\sigma$ level for background test, whereas they lie in the interval $(-0.0143,\; 0.0222)$ \& $(0.0187,\;0.3909)$ for the combined tests.\\

 The best-fit values for $A_{s}$, $B$ and $\alpha$ values estimated for background tests are all positive. But in the case of combined test a negative $\alpha$ with positive $A_{s}$ is allowed for $99.7 \%$ confidence limit. It is also evident from tables-(\ref{tab7}) and (\ref{tab7a}) that a negative $B$ is also permitted but for a physically viable model we consider $B\geq 0$. 
The range and the best-fit values  for $A_{s}$ is positive in the model.\\
 
\begin{figure}
\centering
{\includegraphics[width=230pt,height=200pt]{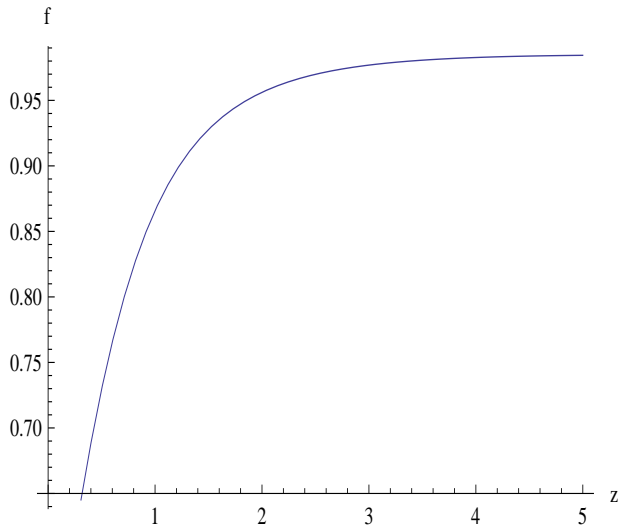}}\\
\caption{Evolution of growth function $f$ with redshifts in MCG}
\label{grf}
\end{figure}

\begin{figure}
\centering
{\includegraphics[width=8cm,height=6cm]{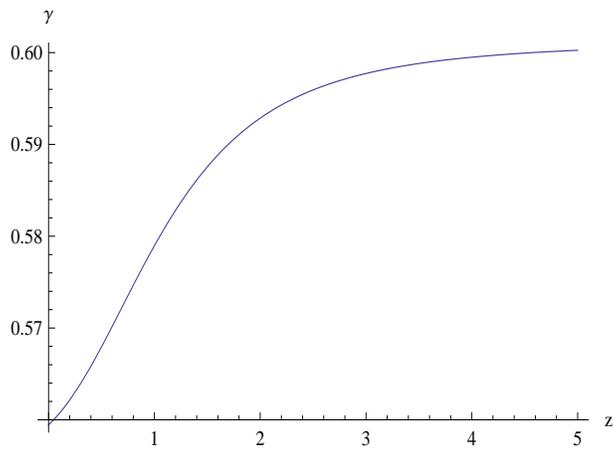}}\\
\caption{Evolution of growth index $\gamma$ with redshifts in MCG }
\label{grind}
\end{figure}

\begin{figure}
\centering
{\includegraphics[width=8cm,height=6cm]{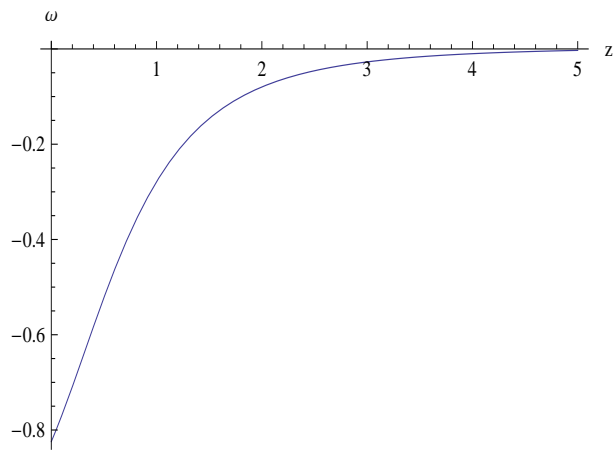}}\\
\caption{Evolution of the state parameter ($\omega$) in MCG}
\label{eos}
\end{figure}

In fig.-\ref{grf} the growth function $f$ is plotted with redshift $z$ using best fit values of model parameters. It is evident that the growth function $f$ lies in the range 0.472 to 1.0 for a variation of redshift from $z=0$ to $z=5$. We note that to begin with $f$ remains a constant but decreases sharply at a lower redshift, indicating that the major growth of our universe occurred in the early epoch at a moderate redshift value. 
In fig.-\ref{grind} the variation of the growth index  $\gamma$  with redshift $z$ is plotted. It is evident that the  growth index  $\gamma$ varies between 0.559 to 0.60 for a variation of  redshift between $z=0$ to $z=5$. A sharp fall in the values  of $\gamma$ at low redshift is evident from the figure.\\
In fig.-\ref{eos} the variation of the equation of  state  $\omega$ is plotted with $z$. It is noted that  $\omega$ varies from  -0.836  at the present epoch  ($z=0$) to $\omega \rightarrow 0$ at an intermediate redshift ($z=5$). This result indicates that the universe is now passing  through an accelerating phase which is dominated by dark energy whereas in the early universe ($z> 5$) it was dominated by matter permitting a decelerating phase.\\ 

\begin{figure}
\centering
{\includegraphics[width=8cm,height=6cm]{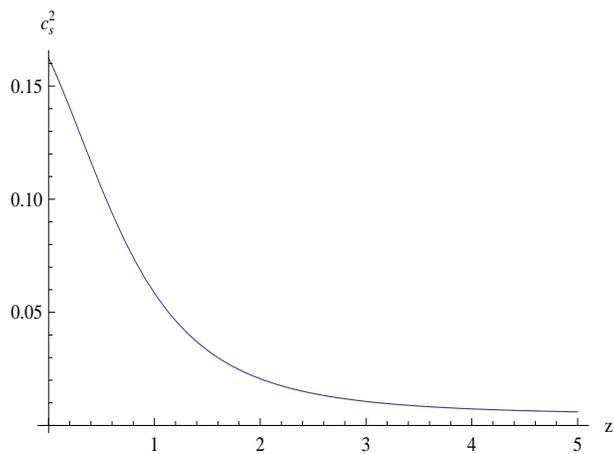}}\\
\caption{Variation of square of the sound speed  in MCG }
\label{spd}
\end{figure}

\begin{figure}
\centering
{\includegraphics[width=8cm,height=6cm]{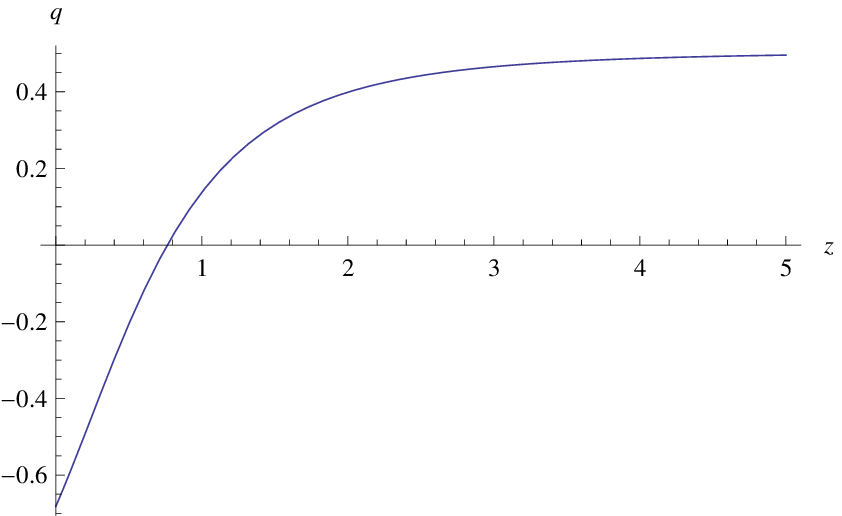}}\\
\caption{Deceleration parameter  with redshift in MCG}
\label{decln}
\end{figure}

The variation of the sound speed $c^2_{s}$ with redshift $z$  is plotted in fig.-\ref{spd}. It is noted that $c^2_{s}$  varies  between 0.162 to 0.01 in the above  redshift range admitting causality. A small positive value indicates the occurrence of growth in the structures of the universe.
The nature of variation of the deceleration parameter is shown in fig-\ref{decln}, which shows that at the present epoch it is $-0.691$ and that there was a decelerating phase in the past as well.

\section{Discussion}

In the paper we have considered dynamical aspects of the universe considering observational data namely, Stern OHD, CMB shift, BAO peak parameter and supernovae. We also considered here two different growth data sets that are relevant for understanding dark energy making use of MCG as a candidate in an FRW universe.
The best-fit values of the parameters $A_{s}$, $B$, $\alpha$ obtained from $\chi^{2}_{tot}(A_{s}$,B,$\alpha)$ are shown in Table-\ref{tab4}.
Using the best fit values, we analyzed the model and determined the allowed range of values of the EoS parameters which  are tabulated in Tables-\ref{tab5},\ref{tab5a},\ref{tab6},\ref{tab6a},\ref{tab7} and \ref{tab7a}.
In the case of combined tests, the range of values of $A_{s}$ and $B$ are found to lie in the intervals $(0.7906,\; 0.8571)$ \& $(-0.0107,\; 0.0203)$,respectively, at 3$\sigma$ level, $A_{s}$ and $\alpha$ lie in the intervals $(0.7678,\; 0.8793)$ \& $(-0.0339,\;0.5048)$, respectively, at 3$\sigma$ level, that of $B$ and $\alpha$ lie in the intervals $(-0.0143,\; 0.0222)$ \& $(0.0187,\;0.3909)$, respectively.
The best-fit values of the growth parameters for the MCG model at the present epoch ($z=0$) is determined which are $f$=0.472, $\gamma$=0.559,  $\omega$=-0.836, $q(0)=-0.691$ and $\Omega_{m0}$=0.261 (shown in Table-\ref{tab8}).
It is also noted that the growth function $f$ varies between 0.472 to 1.0  and the growth index  $\gamma$  varies between 0.559 to 0.60 for a variation of redshift from $z=0$ to $z=5$.
 In this case the equation of state  $\omega$ lies between -0.836 to 0 where the sound speed $c_{s}^2$  varies between  0.162 to 0.01. \\

Thus  a satisfactory cosmological model is permitted accommodating the present accelerating phase of the universe with MCG in GTR-framework. We note that the range of values for the EoS parameters are considerably different from that obtained in the earlier  model \cite{paul13} when CMB shift,BAO peak parameter and supernovae data were considered.
The negative values of the equation of  state  ($\omega \leq$ -1/3) signifies the existence of such a phase of the universe accommodating acceleration. The sound speed obtained in the model is found small which permits structure formation of the universe. Thus the MCG model may be considered as a good candidate for describing  the evolution of the universe which  reproduces the  cosmic growth with inhomogeneity in addition to  a late time accelerating phase.   

In   Table-\ref{tab8}  we present  values of the EoS parameters, present growth parameters and density parameter ($\Omega$)  for both the MCG and GCG models for comparison. The MCG model is considered as a good fit model  with the recent cosmological observations which accommodates recent accelerating phase followed by a decelerating phase. 
In   Table-\ref{tab9}, we present values of the EoS parameters corresponding to the previously studied  GCG model and MCG  models obtained by us. Considering the growth data along with other cosmological observed data, we note that  $\alpha$ remains positive number for almost the entire range (Tables-\ref{tab6},\ref{tab6a},\ref{tab7},\ref{tab7a}). The  best fit value of the parameter $\alpha$  are positive both for the MCG and GCG models. It is evident that the observational constraints on $\omega$, deceleration parameter and  square of sound speed are found considerably higher compared to the previous analysis without the growth test data which indicate structure formation and higher acceleration. Also, it is evident that the present matter contribution is lower (consequently the dark energy contribution is higher) in the current analysis.\\

\begin{figure}
\centering
{\includegraphics[width=8cm,height=6cm]{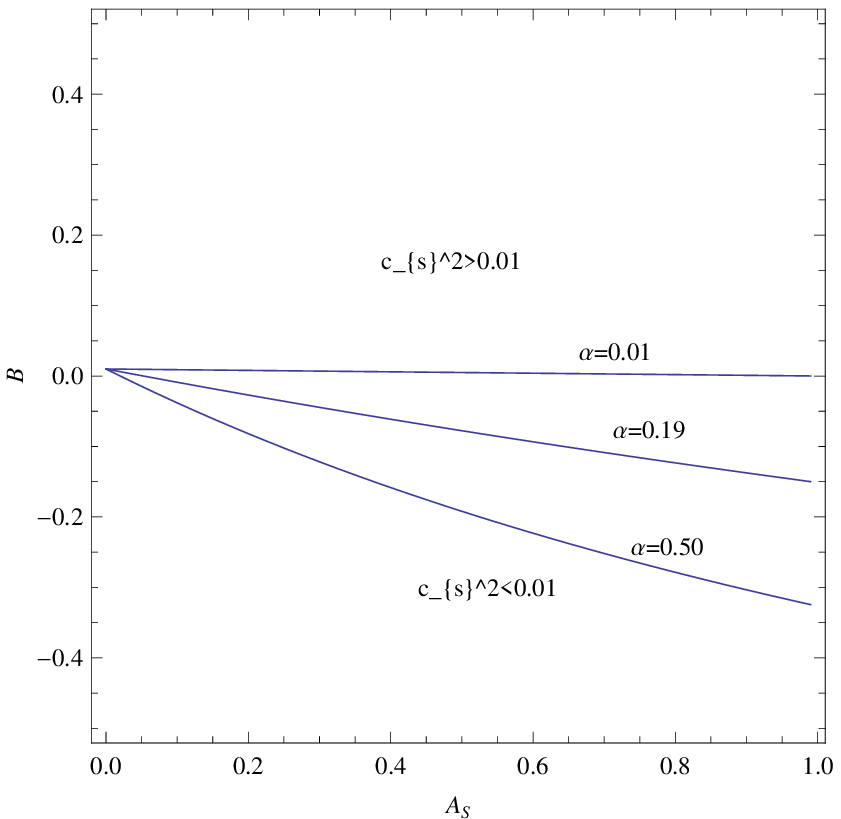}}\\
{\includegraphics[width=8cm,height=5cm]{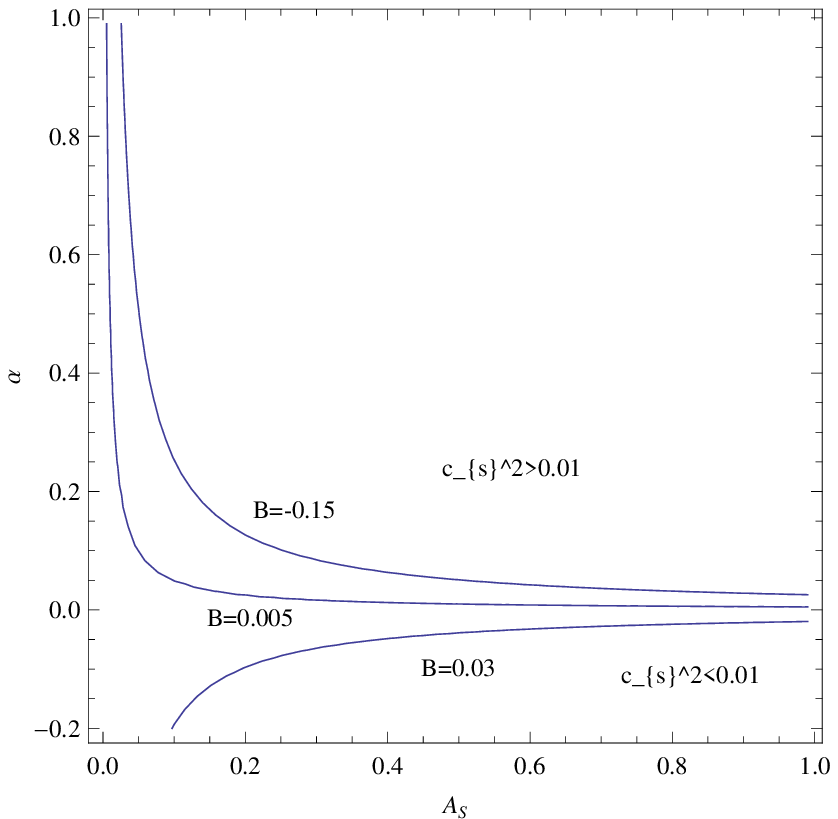}}\\
{\includegraphics[width=8cm,height=5cm]{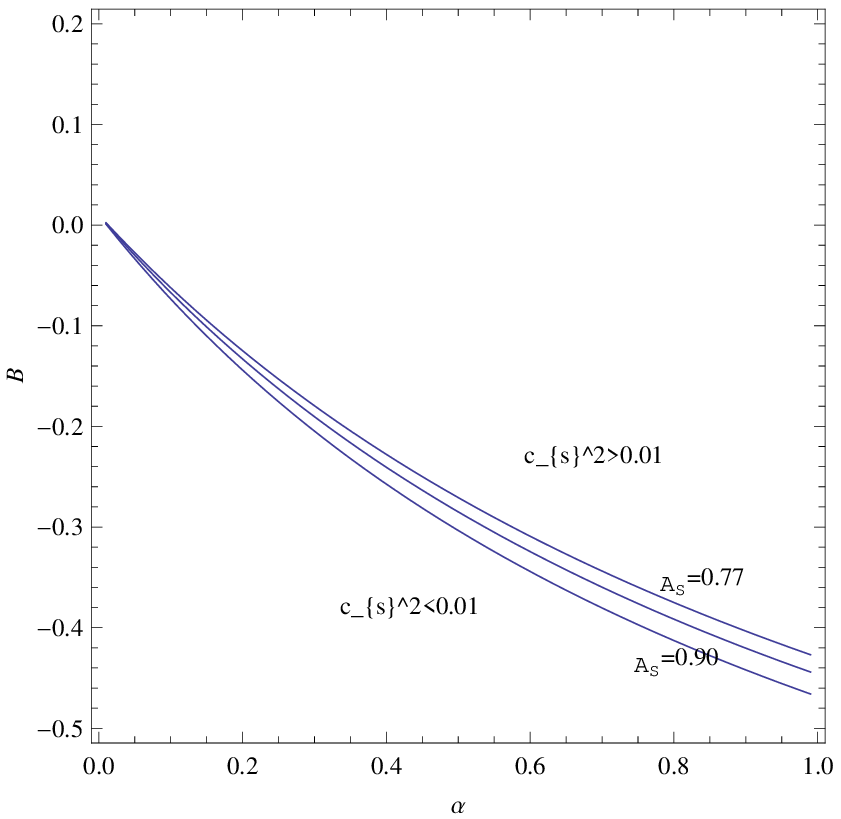}}\\
\caption{Contours for the adiabatic sound speed}
\label{adspd}
\end{figure}

\begin{figure}
\centering
{\includegraphics[width=8cm,height=6cm]{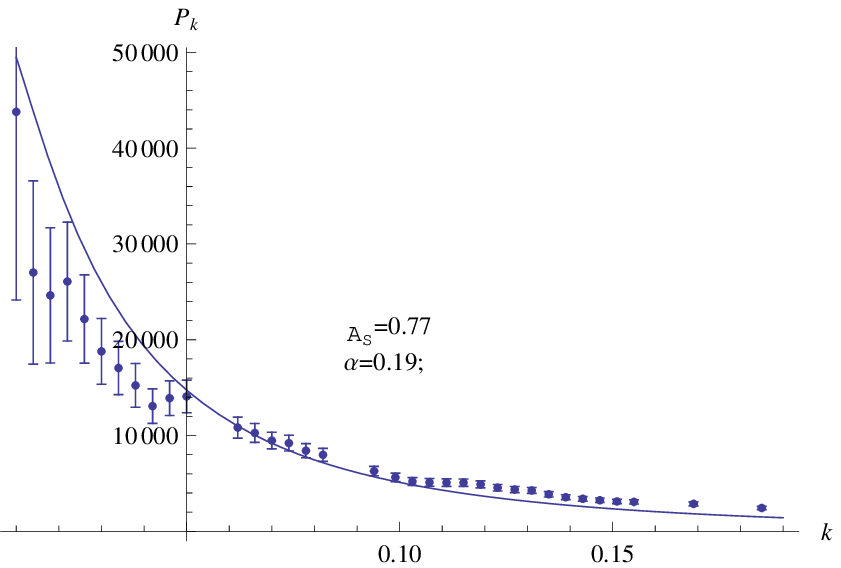}}\\
{\includegraphics[width=8cm,height=5cm]{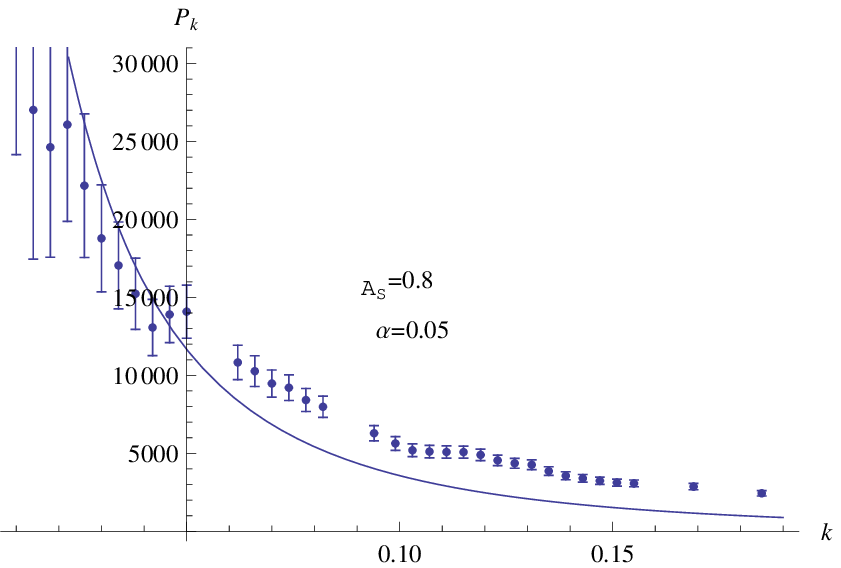}}\\
{\includegraphics[width=8cm,height=5cm]{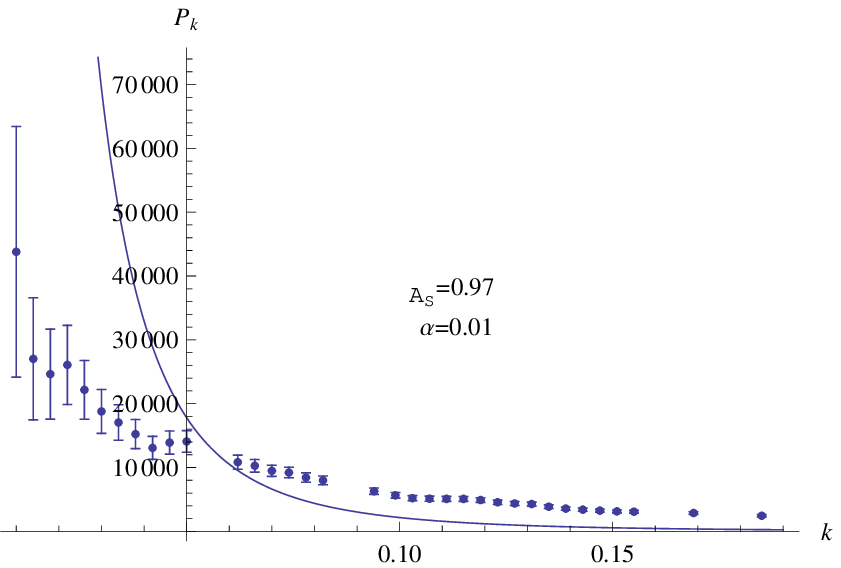}}\\
\caption{power spectrum in MCG}
\label{ps}
\end{figure}

\begin{figure}
\centering
{\includegraphics[width=8cm,height=6cm]{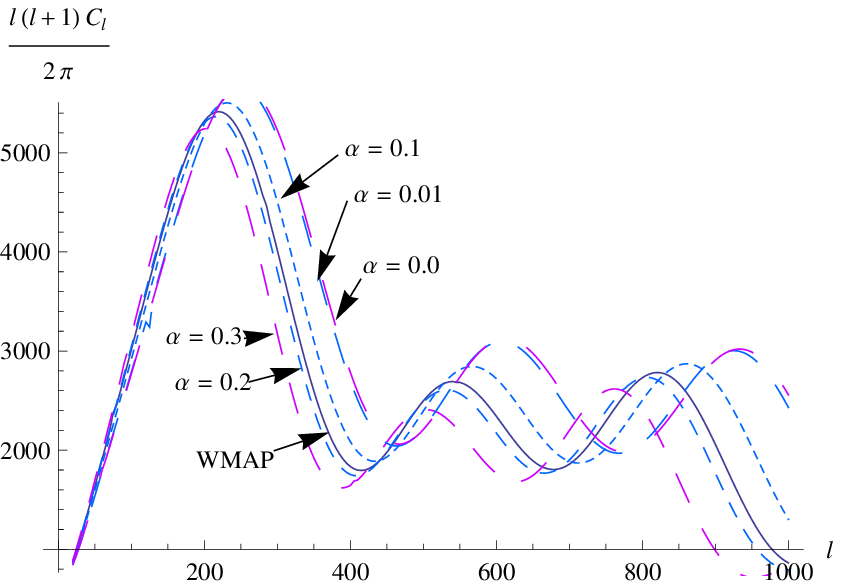}}
\caption{Comparison of the theory with WMAP observations for CMB power spectrum. The solid line corresponds to 3-year WMAP data and  dotted lines correspond to $\alpha$ = $(0, 0.01,0.1,0.2, 0.3)$ of MCG.}
\label{cmb-ps}
\end{figure}

{\bf It is observed from fig \ref{adspd} that the model considered here admits an accelerating universe for a positive B  ($B\approx 0.005$, a best fit value taken from table-\ref{tab4}). It is evident from table \ref{tab8} that the best fit value of the EoS parameter  with MCG ($\omega=-0.84$) lies in the accepted range of values ($-1\leq \omega\leq-0.8$) accommodating accelerating universe.
It is noted that a  small $c_{s}^2$ may be obtained  in the model (fig \ref{adspd}) with lower limiting values of $A_{s}$ and $\alpha$ at $99.7 \%$  confidence limit taken from tables-(\ref{tab6a}, \ref{tab7a}).  
 For a negative $B$ value ($\approx-0.01$ taken from  table-\ref{tab5a}) a very small value of the sound speed ($\approx 10^{-7}$)  which is necessary for adiabatic UDM model is permitted for limiting values of $A_{s}$ and $\alpha$ at $99.7 \%$  confidence limit (table-\ref{tab6a},\ref{tab7a}). This UDM can cluster at linear scale.
We note that the case with $B<0$ and $A<0$ leads to instability which  does not arise in our model as $A=A_{s}(1+B)\rho^{(\alpha+1)}_{0}$ is always positive.  It comes out from the analysis that
MCG is acceptable which may be used to describe satisfactorily the background tests in cosmology.  The shape of the power spectrum in the model obtained from perturbative analysis is shown in fig. (\ref{ps})  for different $A_s$ and $\alpha$. It is evident that for a higher value of $\alpha$ and lower $A_s$ the spectrum fits well with observational data \cite{cole05} in MCG.
 In the MCG model the positivity of the squared sound speed requires both the parameters $B$ and $\alpha$ to be  positive definite. It is also noted that a cosmological model with a small
range of negative values of B and $\alpha$ can in principle are allowed.   
\\
The power spectrum for MCG with observational data \cite{cole05} is plotted in fig. (\ref{ps}) for various $\alpha$ values to study the stability of the model. Unlike GCG \cite{sandvik}   no  oscillations  in the power spectrum for non-zero $\alpha$ in MCG  is observed.
	In  fig -\ref{cmb-ps} we plot the multipole co-efficient $l(l+1)C_{l}/2\pi$ in square microKelvin with  l in the case  of  CMB power spectrum. The solid line drawn corresponds to 3-year WMAP data and different dotted lines are plotted for various values of $\alpha$ $(0, 0.01,0.1,0.2, 0.3)$ of EoS for MCG. It is evident that for the lower values of $\alpha$ the peak rises and shifts towards the higher l value. It is noted that the CMB power spectrum curve for  $\alpha \sim 0.2 $ (near the best-fit value of our model), almost matches with the  WMAP curve. It is also noted that as $\alpha \rightarrow 0$, the peaks of the power spectrum shifts towards higher l values. Thus a non-zero value of $\alpha$ with MCG describes the observational  cosmology. Thus, cosmological model with MCG having non-zero $\alpha$  is found stable unlike GCG where it is unstable. }

\section{Acknowledgements}
The authors would like to thank the {\it IUCAA Reference Centre} at North Bengal University for extending necessary research facilities to initiate the work. BCP would like to thank the {\bf Institute of Mathematical Sciences, Chennai} for hospitality during a visit and TWAS-UNESCO for awarding Associateship.  BCP acknowledges financial support (Major Research Project, Grant No. F. 42-783/2013(SR)) from the University Grants Commission (UGC), New Delhi. PT acknowledges the UGC, New Delhi for financial suport under a Minor Research project (No.F.PSW-073/12-13 (ERO) dated 18.02.13).

\end{document}